\documentclass[aip,jcp,floatfix,reprint]{revtex4-1}
\usepackage[version=3]{mhchem}
\usepackage{amsmath}
\usepackage{amsfonts}
\usepackage{amssymb}
\usepackage{graphicx} 
\usepackage{threeparttable}
\usepackage{booktabs}
\usepackage{siunitx}

\def\mys#1{{\mbox{\tiny{#1}}}}    

\def\collrad{a}                  

\def\etac{\eta_{\mys{c}}}        
\def\F{F}				 		 
\def\FD{F^{\mys{DLVO}}}		 	 
\def\poly{C_{\mys{v}}}           
\def\W{W_{\mys{IM}}}             

\def\kB{k}            			
\def\kBT{\kB T}                 

\def\Deff{D_{\mys{eff}}}		
\def\epr{\epsilon_{\mys{r}}}	
\def\lB{\lambda_\mys{B}}    		
\def\invD{\kappa_{\mys{D}}}		
\def\kap{\kappa}				
\def\pot{U}	 					
\def\DLVO{U^{\mys{DLVO}}}	    

\def\nonadd{\Delta_{\mys{y}}}				
\def\funct{\Delta}
\def\nonDH{\Delta_{\mys{DH}}}	
\def\rr{r}						
\def\rhoion{\rho_{\mys{ion}}}	
\def\point{Q}					
\def\11{S$_{1}$S$_{1}$}    	
\def\22{S$_{2}$S$_{2}$}    	

\begin{document}

\title{Non-additivity of pair interactions in  charged colloids} 
\footnotetext{\textit{$^{a}$School of Chemistry, University of Bristol, Bristol BS8 1TS.}}
\author{Samuel D. Finlayson}
\affiliation{School of Chemistry, University of Bristol, Bristol BS8 1TS, UK.}

\author{Paul Bartlett}
\email[]{p.bartlett@bristol.ac.uk}
\affiliation{School of Chemistry, University of Bristol, Bristol BS8 1TS, UK.}

\date{\today}

\begin{abstract}

It is general wisdom that the pair potential of charged colloids in a liquid may be closely approximated by a Yukawa interaction, as predicted by the classical Derjaguin-Landau-Verwey-Overbeek (DLVO) theory. We experimentally determine the effective forces in a binary mixture of like-charged particles, of species 1 and 2, with blinking optical tweezers.  The measured forces are consistent with a Yukawa pair potential but the (12) cross-interaction is not equal to the geometric mean of the (11) and (22) like-interactions, as expected from DLVO. The deviation is a function of the electrostatic screening length and the size ratio, with the cross-interaction measured being consistently weaker than DLVO predictions. The corresponding non-additivity parameter is negative and grows in magnitude with increased size asymmetry.
\end{abstract}


\maketitle 

\section{Introduction} \label{secintro}

Interactions between charged micro- and nano-scopic particles in fluids play a pivotal role in soft matter\cite{Walker2011}. As many nanoscale materials contain different-sized particles it is crucial to know how the interactions between charged particles in a mixture can  be determined from the properties of the pure species. In the case of binary interactions between point charges in a homogeneous medium, the force is simply proportional to the product of the charges so the pair potential $\pot_{12}$ between two different point charges is a geometric mean of the two like-charge interactions,
\begin{equation}\label{eqideal}
\pot_{12}(\rr) = \sqrt{\pot_{11}(\rr) \pot_{22}(\rr)},
\end{equation}
when all charges have the same sign. This expression is an example of the widely used ideal { Berthelot} mixing rule\cite{Hansen1986}. While Coulombic forces are in this sense ideal, effective interactions operate between charged colloids in a fluid and effective forces will not necessarily obey this mixing rule\cite{Louis2001}. The deviations can be quantified by a so-called non-additivity function $\funct(\rr)$, defined through the relation
\begin{equation}
\funct(\rr) = \frac{[\pot_{12}(\rr)]^{2}}{\pot_{11}(\rr) \pot_{22}(\rr)} - 1.
\end{equation}
For charged suspensions, the interactions between two particles $i$ and $j$ ($1 \le i, j \le 2$) typically have a Yukawa dependence on the separation $\rr$,
\begin{equation} \label{eqUYukawa}
\frac{\pot_{ij}(\rr)}{ \kBT} =  \point_{ij}^{2} \lB  \frac{\exp(-\invD \rr )}{\rr},
\end{equation}
where $\point_{i}=\point_{ii}$ is the Yukawa point charge\cite{Boon2015} of particle $i$, $\lB = e^{2}/(4 \pi \epsilon_{0} \epr \kBT )$ is the Bjerrum length, $\invD = \sqrt{4 \pi \lB \rhoion}$  is the inverse Debye-H\"{u}ckel length, $\epr$ is the relative permittivity of the solvent, and $\rhoion$ is the  overall number density of ions in solution with charges $\pm e$. For Yukawa interactions, the function $\funct(\rr)$ reduces to a single parameter  
\begin{equation}\label{eqyukawanonadditivity}
\nonadd = \frac{\point_{12}^{2}}{\point_{1} \point_{2}} - 1,
\end{equation}
which characterizes the degree of non-additivity\cite{Hopkins2006}. The non-additivity parameter $\nonadd$ can in general  be either positive or negative. Recent primitive model simulations\cite{Allahyarov2009} of the mean force between charged colloids have identified surprisingly large non-zero values of non-additivity, although to-date experimental evidence for non-additivity in charged systems is absent. Accordingly, an important open question is to what extent are the pair interactions between charged micro- and nano-particles in real systems non-additive?

The Deryaguin-Landau-Verwey-Overbeek (DLVO) theory, developed almost 70 years ago\cite{Derjaguin1940,Verwey1948}, has been used extensively to calculate the effective pair potential between charged colloids immersed in a solution of ions. The potential contains three elements: a hard core excluded volume interaction preventing overlap of the particles, attractive terms arising from dispersion forces, and an electrostatic screened Yukawa pair potential which was derived by linearizing the Poisson-Boltzmann (PB) equation in the Debye-H\"{u}ckel (DH) limit.  The repulsive part of the potential between two colloids $i$ and $j$  at a pair separation $\rr$ is assumed to be
\begin{equation}\label{eqDLVObinary}
	\frac{\DLVO_{ij}(\rr)}{\kBT} = 
		\begin{cases} 
				(Z_{i} \vartheta_{i})  (Z_{j} \vartheta_{j})  \lB  \frac{e^{-\invD \rr} }{\rr}, & \text{for } {\rr \geq \collrad_{i} + \collrad_{j}} \\
				\infty, & \text{for } {\rr <\collrad_{i} + \collrad_{j}},
		\end{cases}
\end{equation}
where $eZ_{i}$ and $\collrad_{i}$ is the surface charge and radius of species $i$, and  $\vartheta_{i} = \exp (\invD \collrad_{i} ) /(1 + \invD \collrad_{i}) $ is the enhancement in the Yukawa point charge\cite{Hansen2000} as the ionic atmosphere is excluded from the hard core of each particle so $\point_{i} = Z_{i}\vartheta_{i}$.
For clarity, we label the large particle as species 1, so the size ratio $\gamma = \collrad_{2}/ \collrad_{1} \le 1$.  Inspection of Eqs.~\ref{eqUYukawa}--\ref{eqDLVObinary} reveals that the non-additivity parameter $\nonadd$ is zero in the DLVO theory, since the surface charge is a constant independent of its neighbours. Indeed inspired by DLVO, it is generally accepted wisdom that non-additivity effects in charged systems are either zero\cite{Walker2011}  or else too small to be significant\cite{Yoshizawa2012}. So additive pair potentials, for instance, have been widely employed to rationalize the rich electrostatic self assembly seen in experiments on binary charged nanoparticle and colloidal systems\cite{Assoud2008,Hynninen2009}.

For \textit{symmetric} (equal-sized) mixtures, the DLVO expression for the repulsive pair potential has, in the main, been amply verified both by direct force measurements\cite{Vondermassen1994,Crocker1994,Crocker1996a,Baumgartl2006,Gutsche2007a,Polin2008,Sainis2008a,Sainis2008,Dreyfus2009,Iacovella2010,ElMasri2011,Koehler2011,MontesRuiz-Cabello2014a} (using techniques such as optical tweezers, video microscopy, or colloidal probe atomic force microscopy) as well as first-principles primitive model computer simulations\cite{DAmico1997,Hansen2000,Cuetos2010,Guerrero-Garcia2011,Turesson2012,Gonzalez-Mozuelos2013}. However surprisingly few direct measurements\cite{Crocker1996a,Ruiz-Cabello2013a} have been made on \textit{asymmetric} mixtures formed from either differently-sized or unequally-charged particles.  Furthermore, the only comprehensive study\cite{Ruiz-Cabello2013a} has focussed on the highly-screened limit ($\invD \collrad_{1} \gg 1$), where the thickness  of the ionic double layer ($1/\invD$) is thin by comparison with the particle radius $\collrad$ and the effect of curvature is expected to be small.  

In this work, we demonstrate that electrostatic interactions in mixtures of asymmetrically-sized colloids are non-additive and we quantify the deviations from classical DLVO theory in the limit $\invD \collrad_{1} \ll 1$. Using blinking optical tweezers (BOT) and digital video microscopy we probe directly \textit{in-situ} the forces acting on spherical poly(ionic-liquid) colloids immersed in a nonpolar electrolyte solution.  Our strategy is to use optical tweezers to select four individual colloidal microspheres - two particles of species 1 and two of species 2 - suspended in the same fluid environment. We then measure the symmetric interactions $\pot_{11}(\rr)$  between a pair of particles of type (11), by temporarily relocating the particles of type 2 far away in a distant pair of traps. 
The positions of the particle pairs are then swapped over and the measurements repeated to determine  $\pot_{22}(\rr)$. Finally, by exchanging one particle from the (11) pair with one from the (22) pair we determine  the asymmetric two-body interactions $\pot_{12}(\rr)$ and so evaluate the non-additivity parameter $\nonadd$ in-situ in a fixed environment.

This paper is organized as follows: In Sec.~\ref{secexp}, we describe the details of our model nonpolar colloidal system and the BOT technique used to measure the  \si{\femto \newton}-level repulsive forces operating between individual pairs of particles. Section~\ref{secequal} describes results for the interactions between equal-sized colloids and shows that the force profiles can be accurately described by constant charge boundary conditions. The primary result of this paper, which is presented in Sec.~\ref{secassymetric}, is the finding of substantial degrees of negative non-additivity in the charge interactions between large and small particles. Finally, in Sec.~\ref{sectheory} using analytic theory, we calculate \textit{a priori}  the non-additivity parameter in the DH limit and show how the sign and magnitude of $\nonadd$ may be adjusted by tuning the size or charge asymmetry of the mixture. 

\section{Experiments and Methods} \label{secexp}

\subsection{Materials}
Highly-charged monodisperse  poly(ionic-liquid) [PIL] colloids with radii of 1.28 $\pm$ 0.02 (L), 0.75 $\pm$ 0.01 (S$_{1}$), and 0.51 $\pm$ 0.01  \si{\micro \metre} (S$_{2}$) were synthesized following the procedures outlined by Hussain et al. \cite{Hussain2013}. The PIL microparticles possess a positive charge due to the  dissociation of  lipophilic  ionic liquid groups on the surface of the particle. The colloids charge spontaneously at room temperature and charge control agents are unnecessary. This has the advantage of considerably simplifying interactions so, as we show below, the microparticles are closely approximated by constant charge boundary conditions. The particles consist of a spherical solid core of polymerized methyl methacrylate, methacrylic acid, and a fraction $\W$ (by weight) of a polymerizable ionic monomer. The charged cores were coated by an $\approx$\SI{10}{nm} thick co\-valently-bound layer of  poly(12-hydroxy stearic acid-co-methyl methacrylate) to provide additional steric stabilization at high electrolyte concentrations. For the purpose of this study, two different lipophilic ionic monomers were chosen with the same anion (tetrakis [3,5-bis-(tri-fluoromethyl) phenyl] borate \- [TFPhB]$^{-}$) and different cations ([IM$_{1}$]$^{+}$ and [IM$_{2}$]$^{+}$), whose molecular structures are summarized in Table~\ref{tblstructure}. 
 
 \begin{table}[h]
 	\centering
 	\caption{Molecular Structures of Nonpolar Electrolytes} 
 	\label{tblstructure}
 	\begin{ruledtabular}
 	\begin{tabular}{m{2cm} m{4cm}}
 	Abbreviation &  Molecular structure  \\
 	\hline
 	\\
 	\ce{[IM1]^{+} [TFPhB]-} &   \includegraphics[trim=0 160 0 30, clip, width=0.25\textwidth]{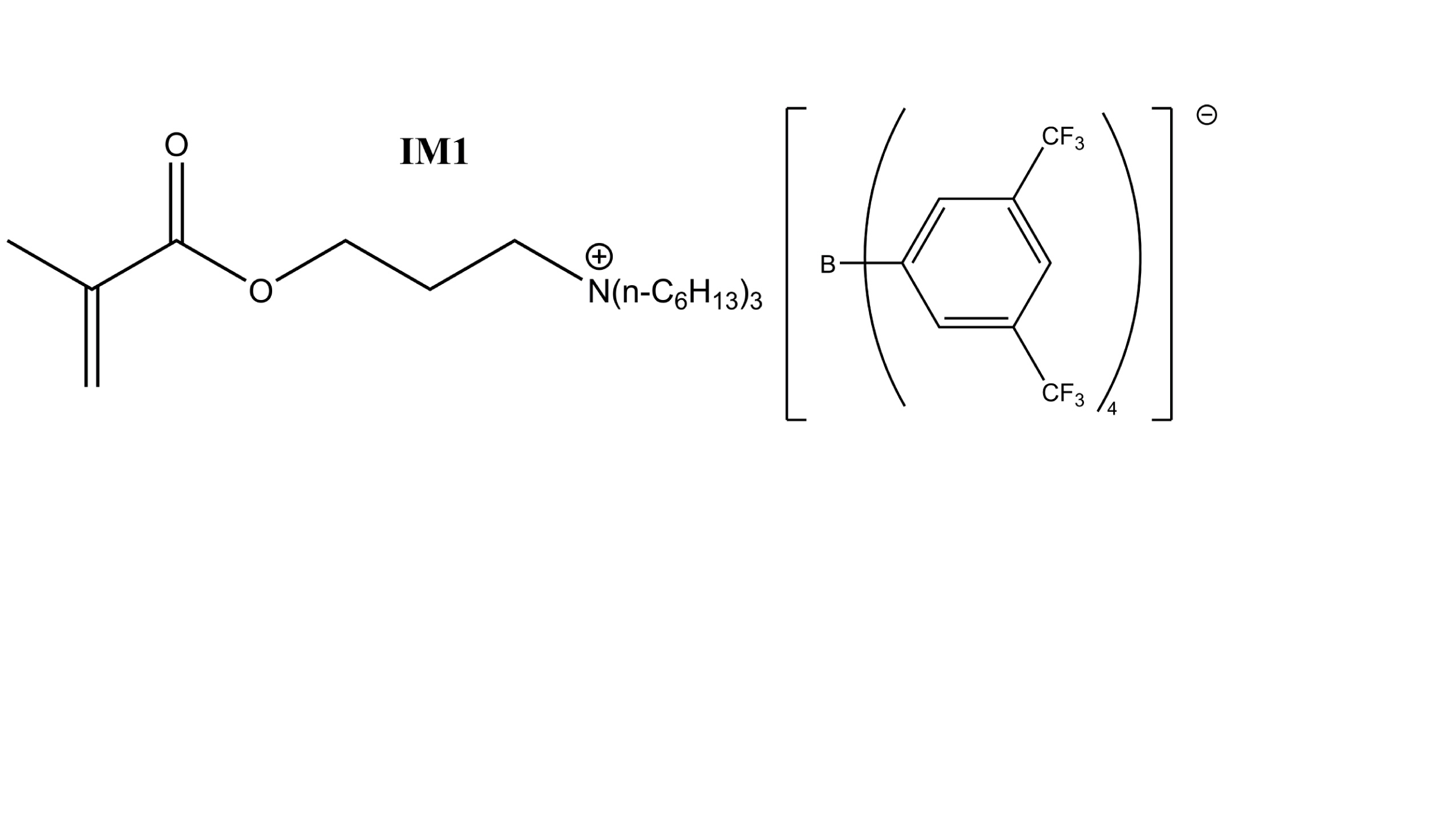}  \\
 	& \\
 	\vspace*{0.1in}
 	\ce{[IM2]^{+} [TFPhB]-} &  \includegraphics[trim=0 40 0 140, clip, width=0.25\textwidth]{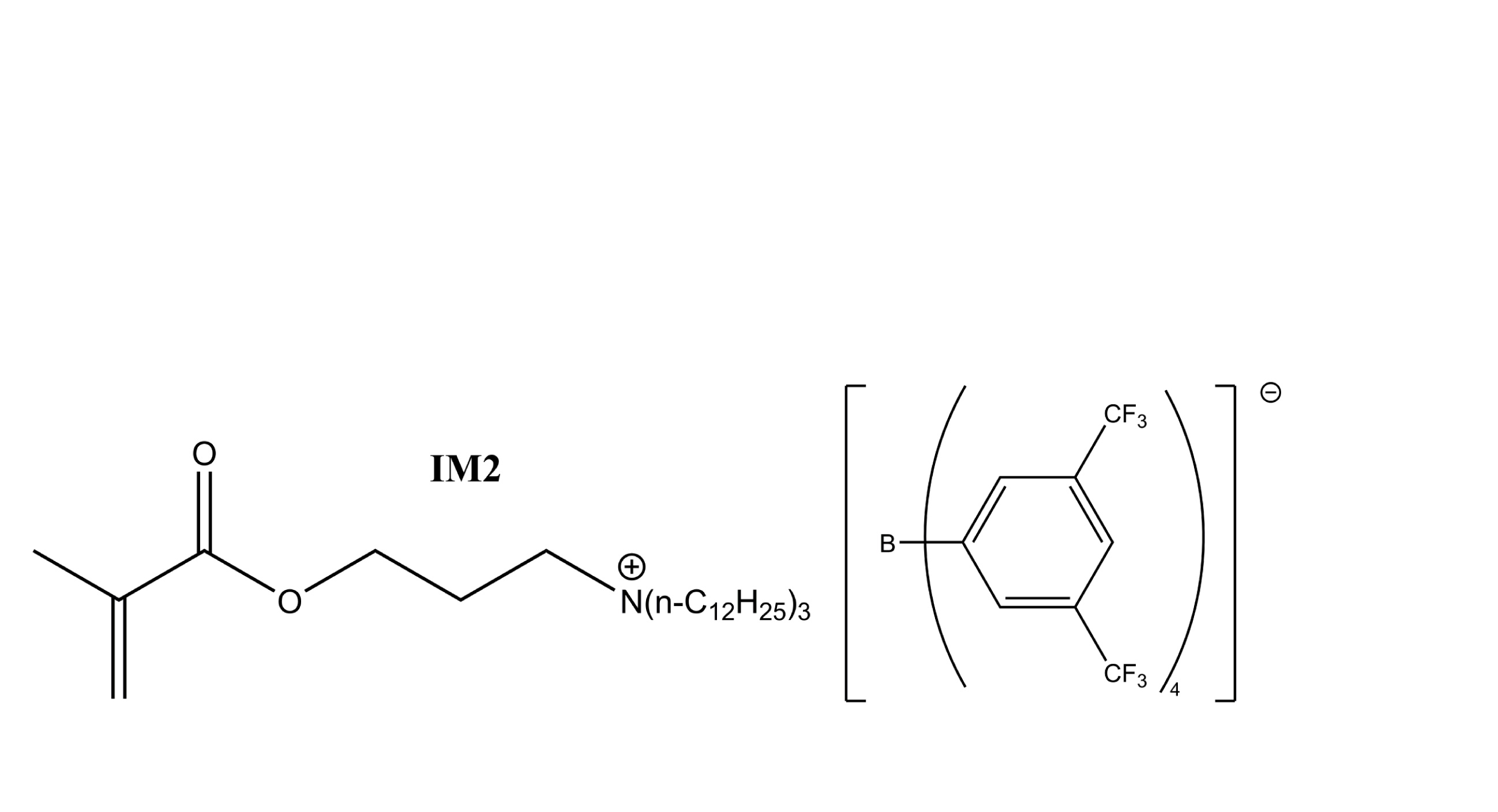} \\
 	\\
 	TDAT &   \includegraphics[trim=00 40 0 40, clip, width=0.2\textwidth]{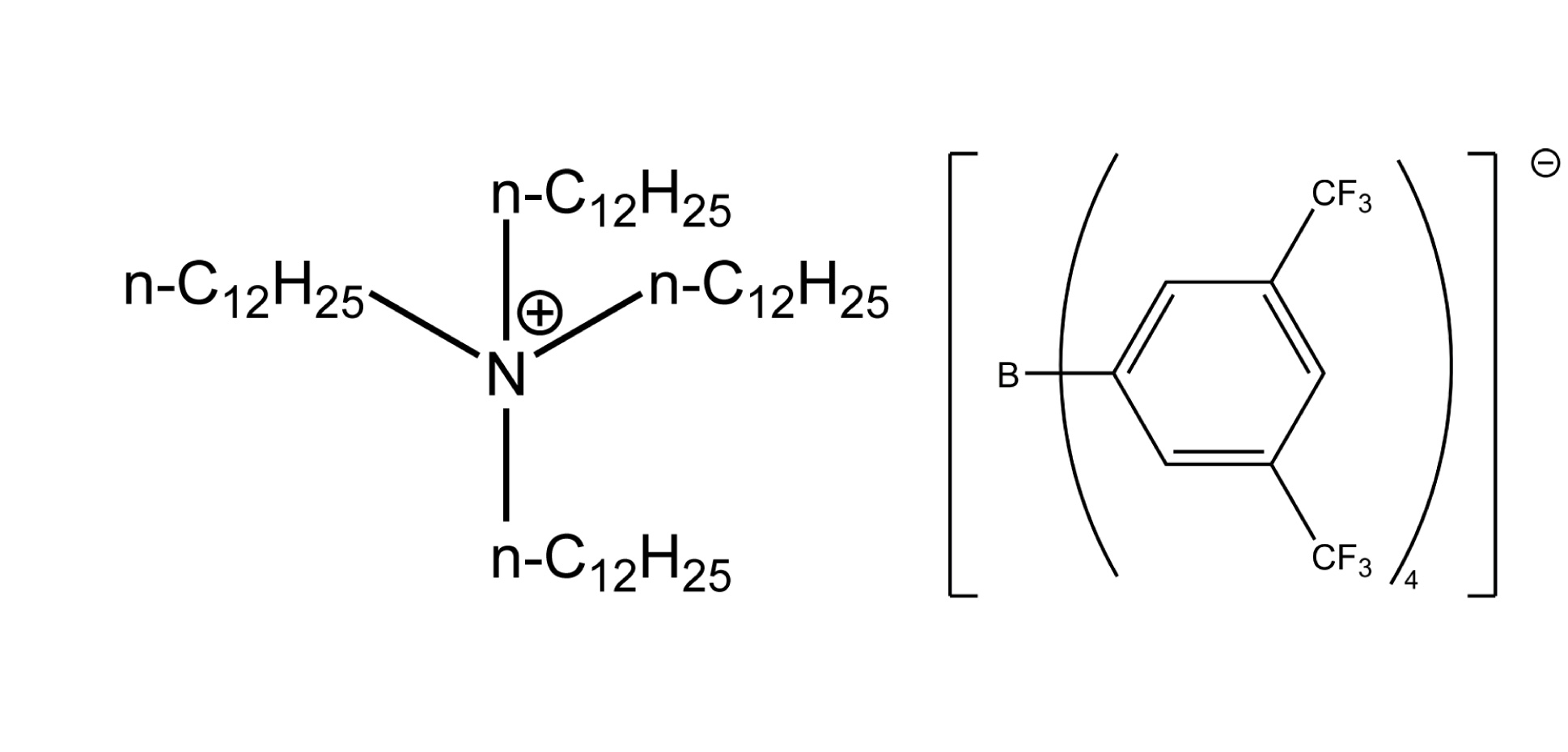}  \\
 	\end{tabular}
 	\end{ruledtabular}
 
 \end{table}

 Dilute suspensions of PIL-microparticles in dry dodecane were used for force measurements. Dodecane was dried with molecular sieves. The ion content of the solvent was monitored by conductivity. Drying continued until the conductivity had fallen to $<$ \SI{1}{\pico \siemens \per \cm}. Microparticles were washed at least 10 times prior to use by centrifugation and redispersal in dry dodecane. A \textit{n}-tetradodecyl ammonium salt [TDAT, Table~\ref{tblstructure}]  was added to the suspensions, at concentrations of up to \SI{80}{\micro \mole  \per \cubic \dm}, to regulate the ionic strength. Table~\ref{tblsamples} compiles details of the three PIL-microparticles used and lists their abbreviations, mean radius $\collrad$, size polydispersity $\poly^{2} = (\left<\collrad^2 \right> -\collrad^{2})/\collrad^{2}$, reduced surface charge $Z \lB / \collrad$, and  composition.

\begin{table*}[]
	\centering
	\caption{Properties of PIL-microparticles used.} 
	\label{tblsamples}
    \begin{ruledtabular}
	\begin{tabular}{cccccccl}
	Abbreviation  & $\collrad$ / \si{\micro \meter} \footnote{Average radius from dynamic light scattering.} & $\poly$ / \% \footnote{Coefficient of radius variation from electron microscopy.} &   $Z \lB / \collrad$ \footnote{Determined from force profiles of symmetric particle pairs.} &  $\W$ / Wt\% \footnote{Weight percentage of ionic monomer as a fraction of total monomer weight.} & Ionic Monomer   \\
	\hline
	L  & 1.28 $\pm$ 0.02 & 9 & $7.4 \pm 0.7$  & 6.0 & \ce{[IM1]^{+} [TFPhB]-}\\
	S$_{1}$  & 0.75 $\pm$ 0.01 & 8 & $5.0 \pm 0.6$  & 8.0 & \ce{[IM1]^{+} [TFPhB]-}  \\
	S$_{2}$  & 0.51 $\pm$ 0.01 & 8 & $5.0 \pm 0.6$ & 2.0 & \ce{[IM2]^{+} [TFPhB]-}  \\
	\end{tabular}
	\end{ruledtabular}
	\vspace{0.5cm}
\end{table*}

\subsection{Force measurements}

The forces between individual pairs of colloidal particles were measured using BOT\cite{Crocker1994,Crocker1996a,Crocker1997,Sainis2007,Merrill2009}, in which optical forces are used to repeatedly  `trap and release' a pair of microparticles. When the optical traps are turned off, the two particles diffuse apart. The interparticle force is inferred from a statistical analysis of the particle trajectories, with a sensitivity of $<$ \SI{1}{\femto \newton}. Before the particles had diffused too far, the laser beams are reapplied so that the particles rapidly return to their initial positions. The trap-and-release cycle is repeated many times to generate high-quality statistics.  The BOT technique is particularly attractive for samples with weak interactions because force measurements occur only when the laser beam is inactive. The optical distortion of the laser fields which occurs near closely-separated particles, and which can lead to deviations between true and measured forces in direct force measurements, is unimportant. In addition, since the sample is not illuminated continuously, there is reduced risk of beam damage and laser heating.

The BOT system was built around an inverted microscope  (Axiovert 200, Carl Zeiss), a high speed digital camera (Dalsa Genie HM640) operating at a frame-rate of \SI{500}{fps}, a 5W  diode-pumped Nd:YAG laser (1064 nm, IPG Photonics), and a LCD spatial light modulator (SLM, PLUTO-NIR, HoloEye)  to generate the holographic traps.  The laser beam was expanded, reflected off the SLM, and then coupled into the back aperture of the microscope objective (Plan Neofluor, 100x, 1.3 NA  Oil, Carl Zeiss).  A beam chopper (Thorlabs MC2000) operating at \SI{20}{Hz} was used to periodically interrupt the laser beam. Dilute dispersions of the PIL-microparticles in dry dodecane were prepared, with colloid volume fractions in the range $1 \times 10^{-5} < \etac < 5 \times 10^{-5}$.   All measurements were conducted approximately \SI{75}{\um} from cell walls to minimize hydrodynamic coupling.  The whole experimental setup was mounted in a temperature controlled 298 $\pm$ 1 \si{\kelvin} room.  To exclude effects arising from slight differences in radii and surface charge, experiments were performed with single pairs of microparticles at each electrolyte concentration. Typically, the forces between the selected particle pair were measured at a range of  separations from \SI{5}{\micro \metre}  to \SI{17}{\micro \metre}. At each separation $5 \times 10^{3}$ complete trap and release cycles were recorded with the particles being allowed to diffuse freely for \SI{25}{\milli \second} before the trapping lasers were re-applied for \SI{25}{\milli \second} to re-establish the initial conditions. During acquisition, $1.5 \times 10^{5}$ digital microscopy images of the two particles were recorded and analysed using custom codes to provide independent measurements of the particle release trajectories $\left(x_{1}(\Delta t),y_{1}(\Delta t) \right)$, and $\left (x_{2}(\Delta t),y_{2}(\Delta t)\right)$ for $0 \leq \Delta t \leq$ \SI{24}{\milli \second}, in \SI{2}{\milli \second} time steps. Tests showed that the tracking algorithms yielded uncertainties in $\rr$ of below \SI{20}{\nano \metre}. Errors in the accuracy of separation measurements due to overlapping particle images, which have been reported previously\cite{Baumgartl2006}, are not expected to be significant here due to the large pair separations used. With these procedures, an experiment with a single pair of microparticles lasted typically \SI{4}{\hour}.

\begin{figure}[h]
%
    \centering
      \includegraphics[width=0.65\textwidth]{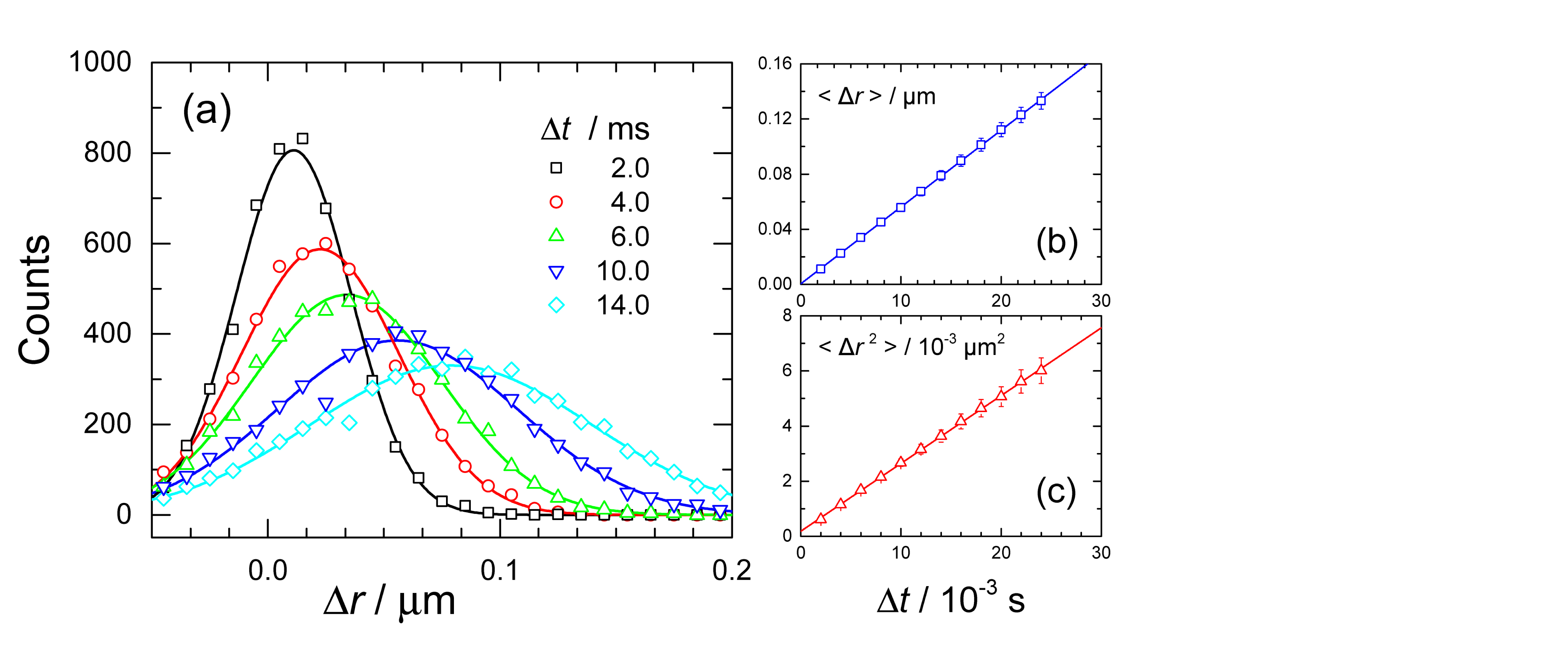}
    \caption{Blinking optical tweezer technique. A single pair of charged particles of radius $\collrad = $ \SI{1.28}{\micro \meter} are held by two holographic optical tweezers (HOT) at an initial separation of $\rr = $ \SI{6.12}{\micro \meter}.   (a) Histogram $P(\Delta \rr)$ of relative pair displacement $\Delta \rr$ in a time interval $\Delta t$, after the HOT laser field is switched off. Solid lines are Gaussian fits to $P(\Delta \rr)$ at fixed $\Delta t$. (b) Mean pair displacement $\left < \Delta \rr \right >$ after time $\Delta t$. The gradient is the drift velocity $v_{\mys{d}} (\rr)$. (c) The mean squared pair displacement $\left < \Delta \rr^{2} \right >$  as a function of elapsed time. The plot is linear showing that the diffusion constant $\Deff(\rr)$ is locally constant over the recorded trajectory. The repulsive force $F(\rr)$ acting between the two particles is $\kBT v_{\mys{d}} (\rr) /\Deff(\rr)$.}
    \label{fgrBOT}
\end{figure}

To determine the force $\F (\rr)$, the probability $P(\Delta \rr, \Delta t)$ of a relative displacement $\Delta \rr$ being observed in a time interval of $\Delta t$  was first calculated from the dataset recorded from a pair of particles at an initial separation of $\rr$. The displacement $\Delta \rr$ at fixed $\Delta t$ arises from an ensemble of Brownian steps so  $P(\Delta \rr)$ is a Gaussian distribution, as illustrated in  Figure~\ref{fgrBOT}(a), with a mean $\left < \Delta \rr \right >$ and a mean-squared displacement $\left < \Delta \rr^{2} \right >$. Because particles diffuse only for a relatively short period ($\Delta t < $\SI{25}{\milli \second}), mean displacements are small in comparison to the initial pair separation ($\left< \Delta \rr \right > / \rr \leq 2\times 10^{-2}$) so we assume that (1) the particles diffuse apart under the action of a fixed force $\F$ and (2) that the diffusion constant $\Deff$ is locally constant over the recorded trajectory. With these assumptions, the gradients of the linear plots of $\left< \Delta \rr \right >$ and $\left< \Delta \rr^{2} \right >$ versus $\Delta t$ (shown in Figure~\ref{fgrBOT}(b) and (c)) provide estimates of the drift velocity $v_{\mys{d}} =  \left< \Delta \rr \right > / \Delta t $ and the local diffusion constant $\Deff = \left< \Delta \rr^{2} \right > / 4 \Delta t$. Following the arguments of \citet{Sainis2007} the pair force is then evaluated from the expression, $\F(\rr) = \kBT v_{\mys{d}}(\rr) / \Deff (\rr)$.

\subsection{Data analysis}

The experimental force-distance profiles were analyzed using a Yukawa model, obtained by differentiating Eq.~\ref{eqUYukawa}
\begin{equation} \label{eqFYukawa}
\frac{\rr^{2}\F_{ij} (\rr)}{\lB \kBT} = \point_{ij}^{2} \exp(-\kap \rr ) \left [ 1 + \kap \rr \right ].
\end{equation}
Here $\rr$ is the center-to-center separation of the two particles, $\kap$ is the inverse screening length (not necessarily equal to $\invD$), and $\point_{ij}$ is an experimentally-determined Yukawa point charge.  Throughout the work detailed below,  the values $T = $ \SI{298}{\kelvin}, $\epr = 2.03$ were used so $\lB =$ \SI{27.6}{\nano \metre} and the force scaling factor ${\lB \kBT}$ is, in convenient units, \SI{0.114}{\femto \newton  \micro \metre \squared}.

\section{Results and Discussion}

\subsection{Symmetric mixtures}\label{secequal}

\begin{figure}[h]
%
%
%
    \centering
      \includegraphics[width=0.4\textwidth]{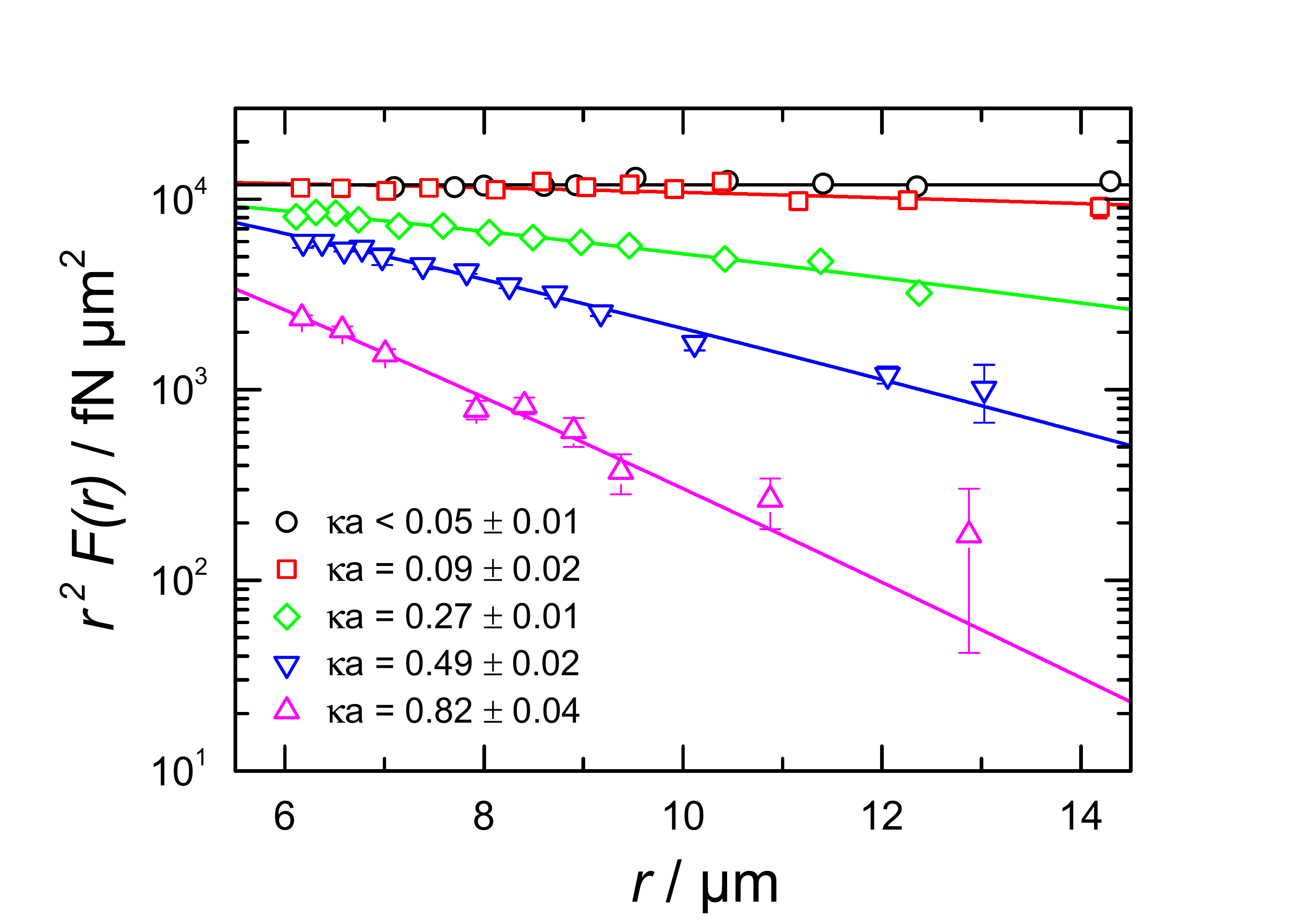}
    \caption{Forces between equal-sized  colloids.  The electrostatic interactions are screened by the addition of the salt \ce{N(Dod)^{+}_{4} [TFPhB]-} (Table~\protect\ref{tblstructure}).  The solid lines depict fits of the experimental data to Eq.~\protect\ref{eqFYukawa}.}
    \label{fgrmeasuredforces}
\end{figure}

We start our analysis of the forces between charged colloids by considering the interactions between identically-sized spheres.   This provides a baseline from which to interpret the more complex interactions of asymmetrically-sized microparticles, as discussed below. The data for $\rr^{2} \F(\rr)$ obtained from BOT measurements on pairs of identical particles  as a function of the center-to-center separation $\rr$, are shown by the symbols in Figure~\ref{fgrmeasuredforces}. Results from five different electrolyte concentrations are depicted:  0.0 (circles), 0.15 (squares), 5.8 (diamonds), 23.0 (down triangles), and 57.0 (up triangles) in units of \si{\micro \mole  \per \cubic \dm} corresponding to $0.05 \le \kap \collrad \le 0.82$. 
In the salt-free limit, the force between particles  is purely Coulombic and $\rr^{2} \F(\rr)$ is a constant, for separations $\rr \leq $ \SI{25}{\micro \meter} where the forces are sufficiently strong to be detectable. With the addition of electrolyte, the interparticle forces display the Yukawa screening regime expected, with the extent of charge screening progressively increasing as the concentration of electrolyte is raised.

\begin{figure}[h]
%
%
%

    \centering
     \includegraphics[width=0.45\textwidth]{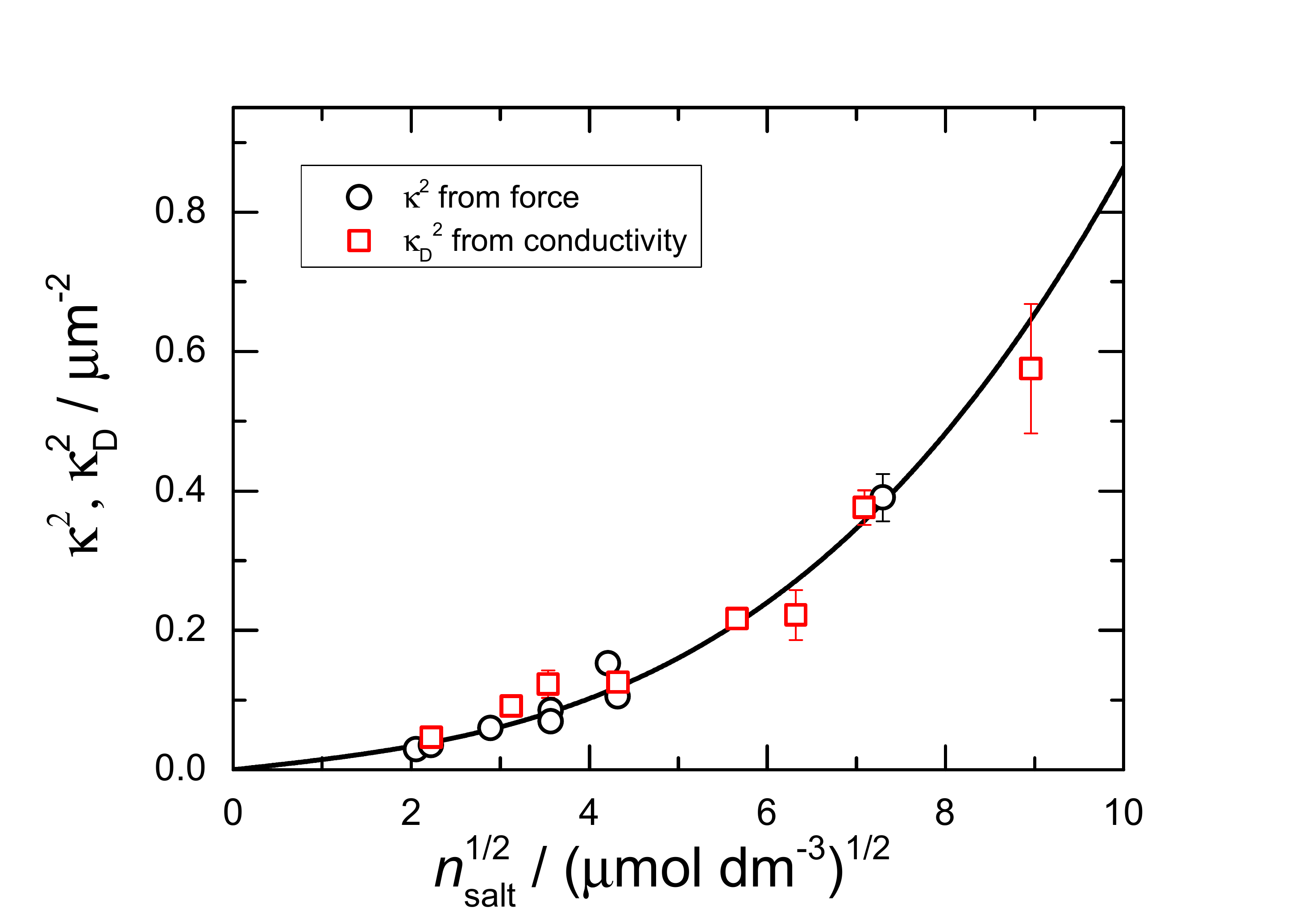}
    \caption{Independent determination of screening length. Comparison of $\kap^{2}$ determined from the measured pair force-distance profiles (black circles) and the Debye-H\"{u}ckel estimate $\invD^{2} = 4 \pi \lB \rhoion$, with $\rhoion$ measured by conductivity (red squares). The solid line is a fit assuming equilibria in solution between free ions, ion pairs, and symmetric triple ions\protect\cite{Fuoss1933}.}
    \label{fgrexpdebye}
\end{figure}

For greater insight into the nature of the screened Coulombic parameters the measured force profiles, at each electrolyte concentration, were fitted to Eq.~\ref{eqFYukawa} (with $i=j$), using $\point_{i}$ and $\kap$ as fit parameters. The solid lines in Fig.~\ref{fgrmeasuredforces} show the resulting fits and demonstrate that the Yukawa expression accurately reproduces the data over the distance range explored. The fitted values of $\kap^{2}$ as a function of the concentration of electrolyte,  together with the DH estimate $\invD^{2} = 4 \pi \lB \rhoion$ with ion concentrations $\rhoion$ measured by conductivity, are shown in Figure~\ref{fgrexpdebye}. While the agreement is not perfect, most of the points agree to within an experimental uncertainty of the interpolated values of the other set { and confirm that the experimental screening  parameter is close to the DH expression.} This is especially striking  since the values for $\kap^{2}$ and $\invD^{2}$ originate from totally independent measurements.

\begin{figure}
%

\centering
     \includegraphics[width=0.45\textwidth]{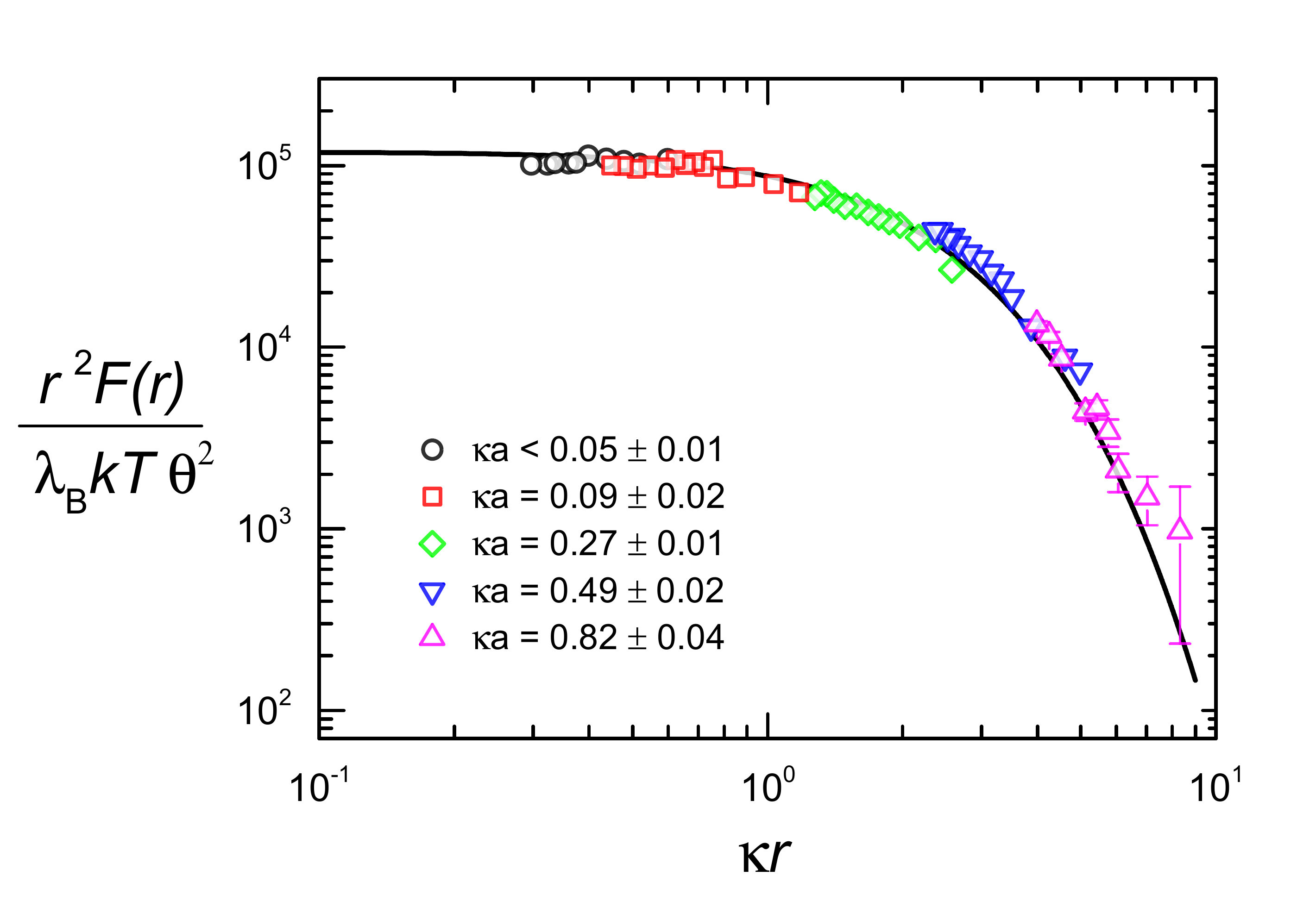}
    \caption{Experimental data from Fig.~\protect\ref{fgrmeasuredforces}, measured at different electrolyte concentrations, replotted as dimensionless force $\rr^{2} F(\rr) /(\lB \kBT \vartheta^{2})$ versus scaled pair separation $\kap \rr$. All experimental data collapse onto a single plot which is closely modelled by the DLVO expression (Eq.~\protect\ref{eqForceDLVO}) with a surface charge (in units of $e$) of $Z = 344 \pm 3$ ($Z \lB / \collrad = 7.44 \pm 0.06$).}
    \label{fgralldata}

\end{figure}

Inspired by the DLVO expression (Eq.~\ref{eqDLVObinary}), we anticipate that if we rescale all the experimental data into the form of $\rr^{2} \F(\rr) / (\lB \kBT \vartheta^{2})$ and replot as a function of $\kap \rr$ then, provided the charge on the particle is a constant, the force profiles measured at different $\kap \collrad$ should all scale onto one single curve. Strikingly, we observe that the data does indeed collapse onto a single master curve (Figure~\ref{fgralldata}).  The DLVO expression for the distance-resolved force 
\begin{equation}\label{eqForceDLVO}
\frac{\rr^{2}\FD (\rr)}{\lB \kBT \vartheta^{2}} = Z^{2} \exp(-\invD \rr ) \left [ 1 + \invD \rr \right ]
\end{equation}
is an excellent fit to the data (the solid line in Fig.~\ref{fgralldata}), confirming the accuracy of the DLVO expression over (almost) two orders of magnitude variation in $\kap \rr$. A surface charge of \SI{344}{\elementarycharge} $\pm$ \SI{3}{\elementarycharge}, independent of the concentration of electrolyte,  is obtained from the fit. 

\begin{figure}[h]
%
    \centering
    	\includegraphics[width=0.45\textwidth]{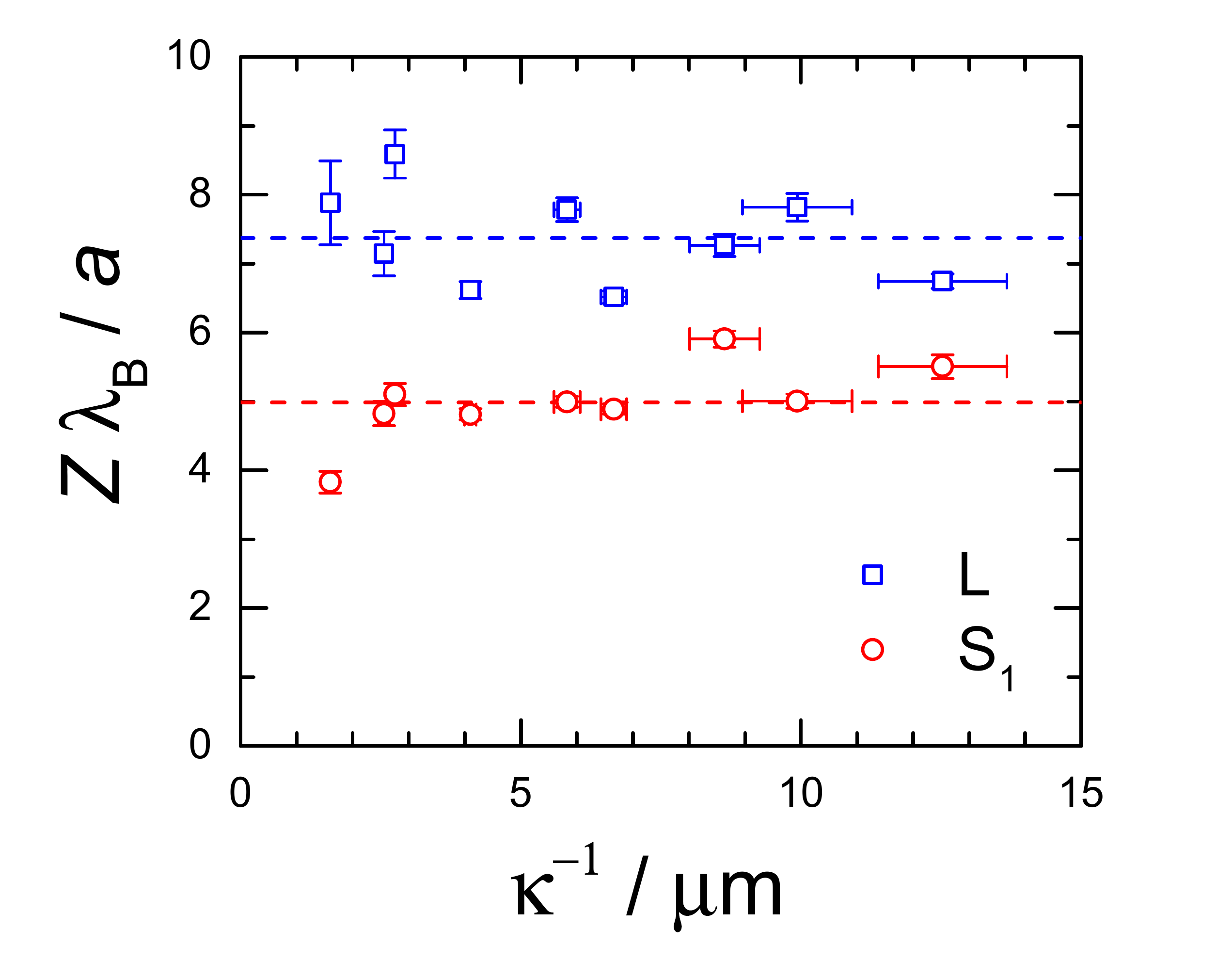} Commented out for submission
    
    \caption{Reduced surface charges $Z \lB / \collrad$ as a function of the (fitted) screening length $\kap^{-1}$.  $Z \lB / \collrad$, within experimental scatter, is independent of the concentration of added electrolyte, which implies that there is no coupling between the ions in solution and the charges on the surface of the particle. Data depicted as squares were obtained from different L--L particle pairs, while the circles denote results from S$_{1}$--S$_{1}$ pairs. The measured $\F(\rr)$ was fitted to the Yukawa expression (Eq.~\protect\ref{eqFYukawa}) using the point charge $\point$ and $\kap$ as adjustable parameters. The particle charge $Z$ was then evaluated from the DLVO prediction, $Z = \point / \vartheta(\kap \collrad )$. The dashed lines correspond to the average reduced charges, reproduced  in Table~\protect\ref{tblsamples}.}
    \label{fgrZeff}
\end{figure}

To confirm that the PIL microparticles are well approximated as a system of hard spheres of constant charge, single pair force measurements were collected on particles of \textit{different} sizes. The results obtained for $Z \lB / \collrad$, the reduced surface charge, as a function of the screening length $1/\kap$ (or equivalently the salt concentration) for LL and \11 particle pairs are depicted in Figure~\ref{fgrZeff} (\22 pairs measured but data not shown). The surface charge $Z$ was evaluated from the DLVO prediction, $Z = Q / \vartheta (\kap \collrad)$, using the Yukawa point charge $Q$ and inverse screening length $\kap$ obtained from a fit to the distance-dependent force profile. The observation, evident in  Fig.~\ref{fgrZeff},  that  $Z \lB / \collrad$ is essentially independent of $1/\kap$ shows that the PIL microparticles behave as a constant charge surface.  This is particularly striking because nonpolar charged systems studied to-date have shown a strong coupling between $Z$ and the concentration of ions in solution\cite{Kanai2015}. The absence of similar effects in the PIL system reflects the strong ionization of ionic-liquid surface groups and the simple surface chemistry.

\subsection{Asymmetric binary mixtures}\label{secassymetric}

To determine the extent of non-additivity  we performed high resolution force measurements on binary suspensions containing a mixture of  microparticles of size ratio $\gamma = \collrad_{2} / \collrad_{1}$, with 1 identifying the large (L) particles, and 2 the small (S) particles.  Holographic optical tweezers were used to move two particles into a region of the sample volume which was devoid of other spheres and at least \SI{75}{\micro \metre} away from the nearest glass wall. The distance-resolved forces were measured in each of the following cases separately: (i) two L spheres, (ii) one L and one S sphere, and finally (iii) two S spheres. The datasets were collected using two pairs of L and S particles to exclude potential complications arising from differences between individual particles. With this procedure a typical experimental run was quite long, lasting some 8-10 \si{\hour}. Careful control of the initial colloid volume fraction was therefore essential to minimize the risk of new particles invading the optical traps. In cases where invasion happened and data acquisition had to be restarted, the measured force curves remained continuous and reproducible (within error), which suggest that the populations of the two species were relatively homogeneous.

\begin{figure}[h] 
%
%
    \centering
       \includegraphics[width=0.35\textwidth]{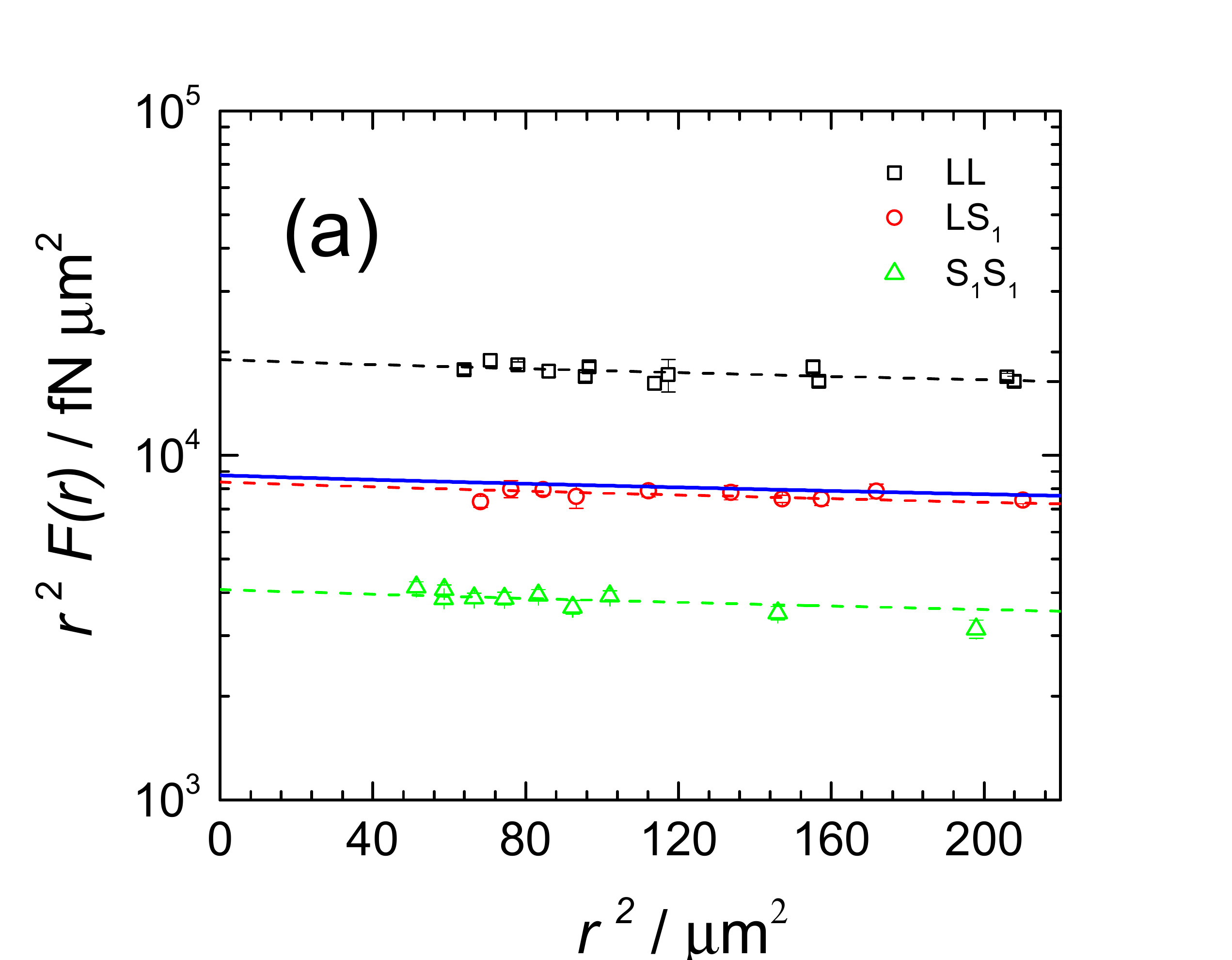}
        \includegraphics[width=0.35\textwidth]{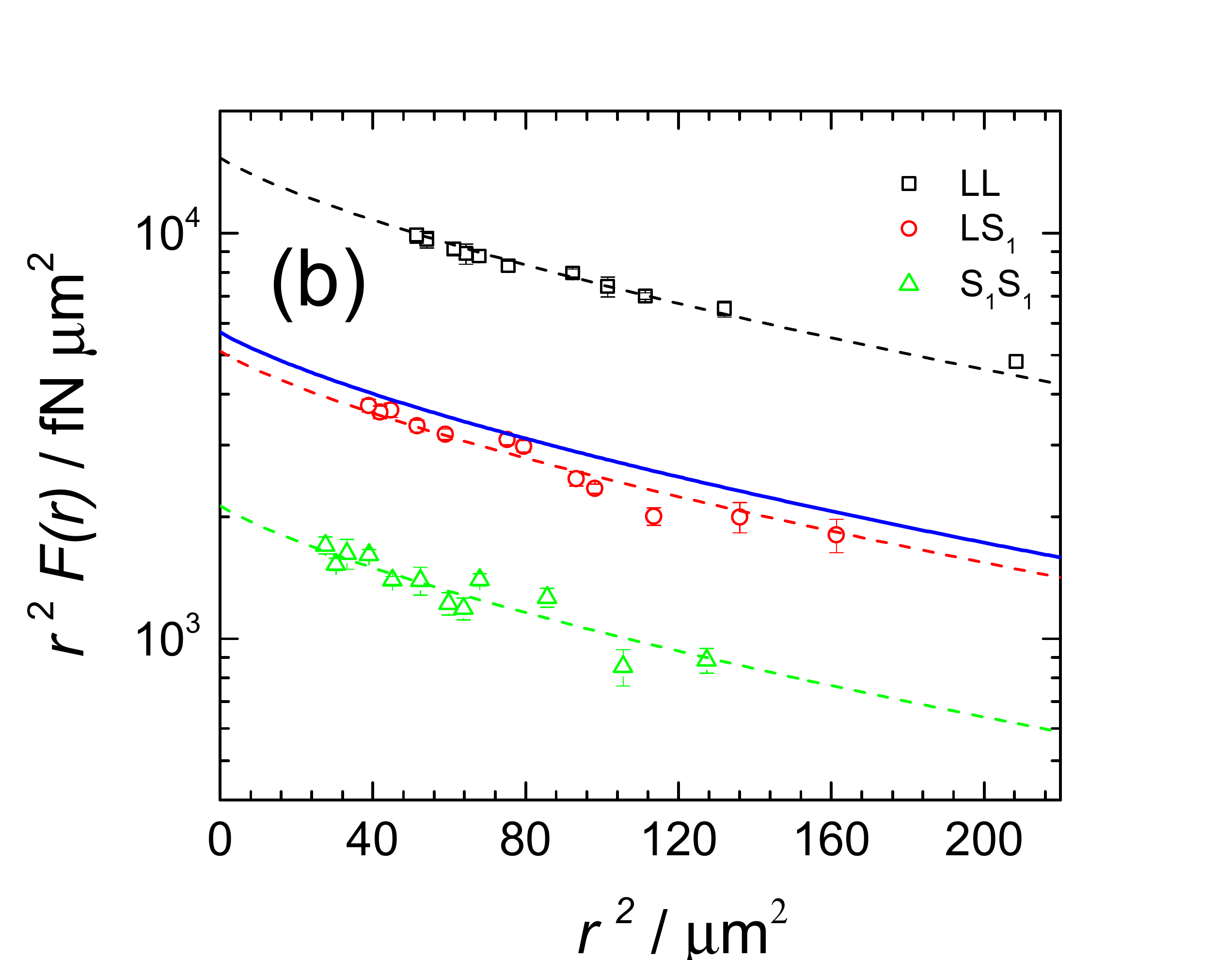}
    \caption{Forces in asymmetric mixtures of radius ratio $\gamma = 0.59 \pm 0.01$. Large spheres (L) have radius $\collrad = $ \SI{1.28}{\micro \meter}, small (S$_{1}$) have $\collrad = $ \SI{0.75}{\micro \meter}. Squares (black) correspond to forces measured between L-L particle pairs, circles (red) to L-S$_{1}$ pairs, and triangles (green) to S$_{1}$-S$_{1}$ pairs. Dashed lines are fits to screened Yukawa expression (Eq.~\protect\ref{eqFYukawa}). Solid lines (blue) show predicted L-S$_{1}$ forces if interactions are assumed additive ($\nonadd = 0$).  For (a) $\kap^{-1} = 23.9 \pm 3.7$ $\mu$m and $\nonadd =-0.011 \pm 0.032$, while for (b) $\kap^{-1} =5.8 \pm 0.2$ $\mu$m  and $\nonadd =-0.10 \pm 0.04$.}
\label{fgrdataLS1}    
\end{figure}

Figure~\ref{fgrdataLS1} shows the scaled pair force $\rr^{2}\F (\rr)$ measured, under exactly the same experimental conditions, in each of the cases (i) [LL], (ii) [LS], and (iii) [SS]. The experiments were performed on mixtures of \SI{1.28}{\micro \metre} and  \SI{0.75}{\micro \metre}  microspheres, with a radius ratio of  $\gamma = 0.59 \pm 0.01$. 
The three force profiles obtained  at each electrolyte concentration were fitted simultaneously to Eq.~\ref{eqFYukawa}. Since all colloids experience the same ionic screening $\kap$ was fixed at the same value for the three datasets. The values of $\kap$ and the three Yukawa point charges $\point_{1}$, $\point_{2}$, and $\point_{12}$, detailing the strengths of the self and cross interactions respectively, were adjusted in a non-linear regression to  fit the experimental data. The fitted force profiles are depicted by the dashed lines  in Fig.~\ref{fgrdataLS1}. 

Classical DLVO theory implies that the strength of the cross-interaction is simply the geometric mean of the self interactions between LL and SS pairs. Specifically, DLVO predicts that the forces between L and S spheres should be
\begin{equation}
\rr^{2}\F^{\mys{DLVO}} /(\lB \kBT) = \point_{1} \point_{2} \exp(-\kap \rr ) \left [ 1 + \kap \rr \right ].
\end{equation}
 Comparison with the experimental data in Fig.~\ref{fgrdataLS1} reveals a significant lack of agreement, with the measured $LS$ forces (circles) being consistently \textit {weaker} than the DLVO predictions (solid lines). The discrepancy is substantial with the force measured between the large and small spheres at, for example $\rr = $ \SI{6}{\micro \meter} and $\kap^{-1} = $ \SI{5.8}{\micro \meter},  some \SI{12}{\femto \newton} less than the force (\SI{104}{\femto \newton}) estimated from classical DLVO, well outside the measurement accuracy of order \SI{1}{\femto \newton}. A close inspection of the data suggests that the deviation is a non-monotonic function of the screening length, with DLVO providing a good approximation to the measured LS interaction in the limit of either strong screening ($\kap^{-1} \rightarrow 0$) or weak screening ($\kap^{-1} \rightarrow \infty$). At intermediate values of $\kap^{-1}$, the discrepancy between experiments and the DLVO predictions is maximized. 

\begin{figure}[h]
%
    \centering
    \includegraphics[width=0.45\textwidth]{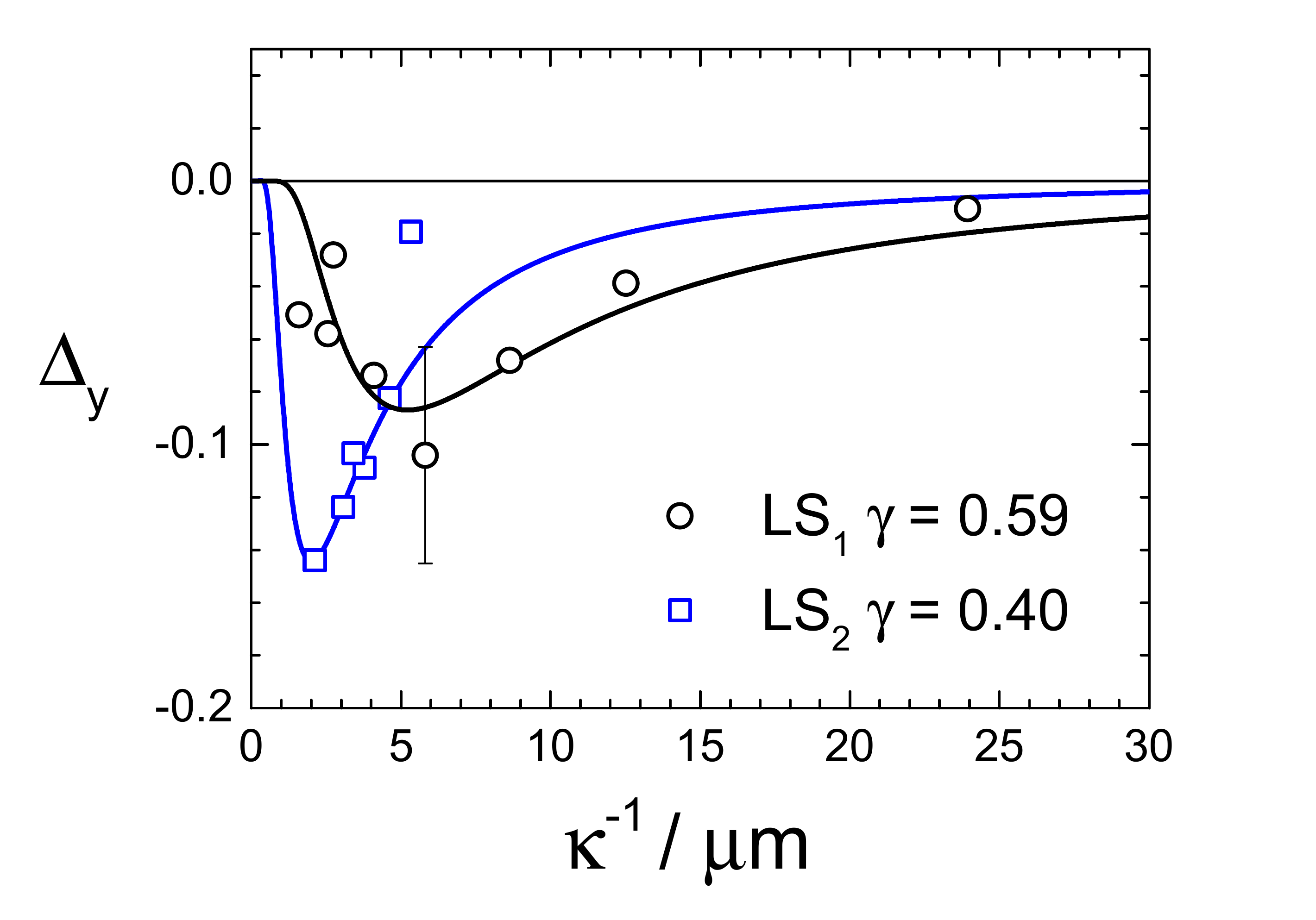}
    \caption{Non-additivity parameter $\nonadd$ for asymmetric-sized Yukawa mixtures, determined from experimental force profiles as a function of the measured screening length $\kap^{-1}$. The lines are scaled Debye-H\"{u}ckel predictions (see text for details).}
    \label{fgrnonaddall}
\end{figure}

The non-additivity parameter $\nonadd$ was extracted from the fitted Yukawa charges, following Eq.~\ref{eqyukawanonadditivity}. The experimental results for the LS$_{1}$ mixture are depicted by the circles in Figure~\ref{fgrnonaddall}. The non-additivity parameter $\nonadd$ is negative at all screening lengths with a minimum of $\approx - 0.08$ at $\kap^{-1} \approx$ \SI{5}{\micro \metre}. To explore the effect of size asymmetry on $\nonadd$,  measurements were performed on a second mixture (LS$_{2}$) with a smaller size ratio of $\gamma=0.40 \pm 0.01$. The results obtained are depicted by the squares in Fig.~\ref{fgrnonaddall}. The data confirms the sign of the non-additivity parameter and imply that the non-additivity \textit{increases} as the particles becomes more asymmetric in size. So, for instance at $\gamma = 0.40$, the depth of the minimum in $\nonadd$ has grown to $\approx - 0.14$.

The substantial non-additivity evident in Fig.~\ref{fgrnonaddall} is, at first sight, rather surprising particularly when DLVO theory seems apparently to do an excellent job reproducing the repulsive forces between same-sized particles, as seen in Fig.~\ref{fgralldata}. Experimentally, there have been  few direct measurements of the non-additivity parameter $\nonadd$. Twenty years ago \citet{Crocker1996a} studied mixtures of asymmetric-sized polystyrene particles in water ($\invD \collrad_{1}\approx 3$) and reported three datasets, two with $\nonadd > 0$ and one with $\nonadd < 0$. The quality of the data was not high so they concluded, on balance, that $\nonadd \approx 0$. More recently, multi-particle colloidal probe techniques\cite{Ruiz-Cabello2013a,MontesRuiz-Cabello2014a} have been used to investigate the forces between charged latex particles with radius ratios down to $\gamma = 0.34$, under conditions of strong screening ($\invD \collrad_{1} \approx 17$). The authors found that under these conditions DLVO theory accounted successfully for the interactions in asymmetric systems suggesting $\nonadd = 0$ in the limit $\invD \collrad_{1} \gg 1$. The first indication that there might be significant deviations from additivity in the weakly-screened limit came from primitive model simulations\cite{Allahyarov2009} of the mean force between two charged colloids, suspended in a bath of explicit ions. \citet{Allahyarov2009} working under conditions of weak screening ($\invD \collrad_{1} = 0.4$), evaluated the forces in asymmetric Yukawa mixtures of size ratio $\gamma = 0.56$ with a constant surface charge of $Z \lB / \collrad_{i} = 6$, and found a surprisingly large and negative non-additivity parameter of $\nonadd = - 0.11$. Our measurements on the LS$_{1}$ mixture were performed under broadly comparable conditions ($\gamma = 0.60$, $\invD \collrad_{1} = 0.26$, $Z \lB / \collrad_{1} = 7.4$, $Z \lB / \collrad_{2} = 5$) to those used in the simulations so the close agreement between the non-additivity parameters from simulation and experiment, is particularly striking and suggests that the mechanism which produces negative non-additivity in these mixtures is robust and general.

\section{Non-additivity in Debye-H\"{u}ckel limit}\label{sectheory}

To account for these observations, we generalize the classical  theory\cite{Derjaguin1940,Verwey1948} of Derjaguin, Landau, Verwey, and Overbeek (DLVO) to incorporate corrections that arise as the double layer around one particle is distorted by overlap with the hard core of a neighbouring particle. To begin, we consider the interaction between two spherical and impenetrable microparticles (denoted $i$ and $j$) a distance $\rr$ apart.
An analytic expression for the electrostatic portion of the interaction at large $\rr$ was obtained first by Derjaguin\cite{Derjaguin1940}, using a linear superposition approximation (LSA). The total electrostatic potential $\Phi$ outside the two particles is equated to the superposition of the potentials ($\Phi_{i}$ and $\Phi_{j}$) produced by each particle in the absence of each other.  The potential distribution outside a particle, if ion-ion correlations effects are neglected, is given by the PB equation, ${\nabla}^{2} \Phi_{i}({ r}_{i}) = \invD^{2} \sinh \Phi_{i}(r_{i})$, where  $\Phi_{i}({ r}_{i}) = e \phi_{i}({ r}_{i}) / \kBT$, $\phi_{i}$ is the electrostatic potential, and $r_{i}$ is the distance to the centre of the particle. For small particle charges, where $|\Phi_{i}| \ll 1$, this equation may be linearized to yield the DH expression for the electrostatic potential, namely
\begin{equation}\label{eqDH}
{\nabla}^{2} \Phi_{i}({ r}_{i}) = 
		\begin{cases}
			0 & \text{for } {r}_{i} < \collrad_{i} \\
			\invD^{2} \Phi_{i}({r}_{i}) & \text{for } {r}_{i} \geq \collrad_{i} .
		\end{cases}
\end{equation}
 In a spherical geometry, the DH equation (\ref{eqDH}) has the analytic solution
 \begin{equation}\label{eqDHpot}
	  \Phi_{i}({ r}_{i})  =
		 \begin{cases}
			  \frac{Z_{i} \lB}{(1+\invD \collrad_{i})\collrad_{i} }  	& \text{for } {r}_{i} < \collrad_{i} \\
			  Z_{i}  \vartheta_{i} \lB  \frac{e^{-\invD \rr_{i}}}{\rr_{i}}        & \text{for } {r}_{i} \geq \collrad_{i} 
			 
		 \end{cases}
 \end{equation}
where  $e Z_{i}$ is the surface charge of particle $i$, and $\vartheta_{i}$ is the charge enhancement factor.  Equation~\ref{eqDHpot} demonstrates that the electrostatic potential $\Phi_{i}$, outside a sphere, is the same as a point charge $\point_{i} = Z_{i} \vartheta_{i}$ located at its centre. So, within the LSA, the pair potential between two charged colloids is identical to the interaction of a pair of point charges $\point_{i} = Z_{i} \vartheta_{i}$ and $\point_{j}=Z_{j} \vartheta_{j}$ immersed in an electrolyte\cite{BellLevineMcCartney1970}, as quoted in Eq.~\ref{eqDLVObinary}.

\begin{figure}[h]
	%
	%
	\centering
	\includegraphics[trim= 270 80 300 20, width=0.3\textwidth]{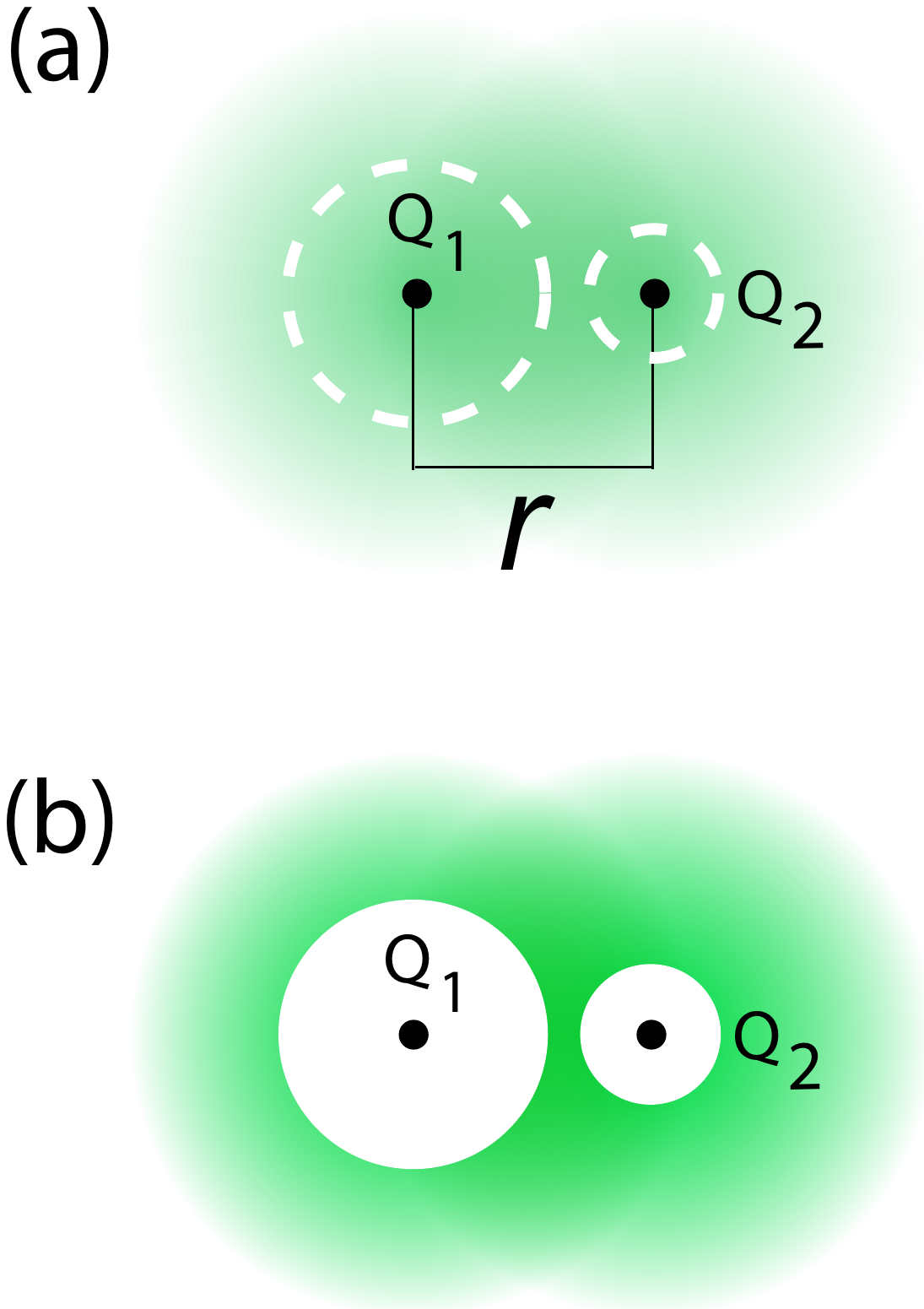}
	\caption{(a) Illustration of the DLVO model, in the weak screening limit ($\invD \rr \lesssim 1$), where the forces may be modelled by a Yukawa interaction between point charges $\point_{1}$ and $\point_{2}$. (b) Exclusion of the diffuse ion cloud (green) from the hard core of each colloid.}
	\label{fgrschematic}
\end{figure}

Although the DLVO pair potential (Eq.~\ref{eqDLVObinary}) has been used extensively, the expression is not exact even in the DH limit. The origin of the breakdown is illustrated graphically in Figure~\ref{fgrschematic}.  Part (a) shows the situation implicit in DLVO, where the density of the ion cloud around the point charges, positioned at the centre of each particle, is sketched. Since the electric field is produced by point charges with no excluded volume and the DH equation is linear, the total electrostatic potential satisfies the LSA, $\Phi = \Phi_{i}+\Phi_{j}$.  In reality, of course, the charges of the ionic sea cannot penetrate the hard cores of the particles centred on each point charge (Fig.~\ref{fgrschematic}(b)). So, as the diffuse layer around one particle starts to inter-penetrate the hard core of a neighbouring particle the LSA will become less accurate. Exclusion of ions from the core of the second particle leads to an increased distortion of the ion cloud as $\rr$ is progressively reduced, which is expected to lead to an enhanced repulsion, in comparison to the DLVO predictions. 

An exact expression for the electrostatic potential $\Phi$, in the vicinity of two identical charged particles, has been long known\cite{Verwey1948} but the solution only exists in the form of an infinite series whose rather complicated nature makes it difficult to truncate and which provides only limited physical insight. The accuracy of the DLVO predictions however may be improved whilst retaining much of its simplicity by using results produced by \citet{Fisher1994} {(FLL)}. Within DH theory, the authors calculate analytic expressions for the interactions between a pair of  ions, treated as charged hard spheres of dielectric constant $\epsilon_{p}$, immersed in an ionic medium of dielectric constant $\epr$. The two-centre nature of the problem makes the analysis demanding. But by using a novel expansion in spherical Bessel functions, { FLL}\cite{Fisher1994} obtained exact, closed-form expressions for the leading and first-order correction term to the interaction free energy $\pot_{ij}(\rr)$, as $\rr \rightarrow \infty$ for all $\invD \geq 0$.

\citet{Fisher1994}  derived the asymptotic expression
\begin{equation}\label{eqfisherii}
\frac{\pot_{ii}(\rr)}{\kBT} = Z_{i}^{2}\vartheta_{i}^{2}\lB \frac{e^{-\invD \rr}}{\rr} + \frac{1}{3} Z_{i}^{2}\vartheta_{i}^{2}\lB \invD^{2}\collrad_{i}^{3}\frac{e^{-2\invD \rr}}{\rr^{2}}
\end{equation}
for the pair potential (without polarization effects, $\epsilon_{p} = \epr$) between two identical charged spheres of radius $\collrad_{i}$ with  $\invD > 0$, in the limit of $\rr \rightarrow \infty$. The first term is the expected DLVO interaction (Eq.~\ref{eqDLVObinary}). The second term, which is the first correction to DLVO, is always positive signifying an enhanced repulsion. The increased repulsion is seen as proportional to the volume of the particle and is strongly screened by a factor of $\exp (-2 \invD \rr) / \rr^{2}$. Both characteristics have a natural  physical explanation. \citet{Li1994} show that the second term in Eq.~\ref{eqfisherii} is simply the free energy to create a cavity of radius $\collrad_{i}$, corresponding to the exclusion of the ionic atmosphere from the hard core of a neighbouring particle as illustrated in Fig.~\ref{fgrschematic}(b). This additional ion exclusion term should scale therefore with the volume of the cavity and to decay as the square of the screened Coulomb coupling, because both the potential at the centre of the neighbouring particle and the amount of charge displaced will vary as $\exp (-\invD \rr) / \rr$.

{
The FLL theory yields an analytical expression for the degree of non-additivity. To that end, it is helpful to express the asymptotic pair potential between charged particles of type $i$ and $j$ as
\begin{equation}\label{eqcorrection}
\pot_{ij}(\rr) =  \DLVO_{ij}(\rr)  \cdot \left[1+\Omega_{ij}(\rr) \right]
\end{equation}
where $\Omega_{ij}$ is the correction to the classical DLVO potential of Eq.~\ref{eqDLVObinary} due to ion exclusion. For identical charged particles, the asymptotic pair potential of Eq.~\ref{eqfisherii} yields
\begin{equation}\label{eqchargecavityii}
\Omega_{ii}(\rr) = \frac{1}{3}(\invD \collrad_{i})^{3}  \frac{e^{-\invD \rr }}{\invD \rr},
\end{equation}
where, as expected, the correction to DLVO is seen to scale with the particle volume and the strength of the screened Coulomb coupling. FLL derived an asymptotic expansion, at the same level of approximation as Eq.~\ref{eqfisherii}, for the  interaction of two \textit{asymmetric} particles, 
\begin{eqnarray}\label{eqfisherij}
\frac{\pot_{ij}(\rr)}{\kBT} & = & Z_{i}Z_{j}\vartheta_{i} \vartheta_{j}\lB \frac{e^{-\invD \rr}}{\rr} + \frac{1}{6} Z_{i}^{2}\vartheta_{i}^{2}\lB \invD^{2}\collrad_{j}^{3}\frac{e^{-2\invD \rr}}{\rr^{2}} \nonumber \\
 &  + & \frac{1}{6} Z_{j}^{2}\vartheta_{j}^{2}\lB \invD^{2}\collrad_{i}^{3}\frac{e^{-2\invD \rr}}{\rr^{2}},
\end{eqnarray}
valid in the limit of $\rr \rightarrow \infty$ and $\invD > 0$.  The correction to the DLVO repulsions between dissimilar particles is then, from Eqs.~\ref{eqcorrection} and \ref{eqfisherij},
\begin{equation}\label{eqchargecavityij}
\Omega_{ij}(\rr) =  \frac{1}{6}  \left [ (\invD \collrad_{i})^{3} \frac{\point_{j}}{\point_{i}} + (\invD \collrad_{j})^{3}  \frac{\point_{i}}{\point_{j}} \right ] \frac{e^{-\invD \rr }}{\invD \rr}.
\end{equation}
Interestingly we note that $\Omega_{ij}(\rr)$, in contrast to Eq.~\ref{eqchargecavityii}, depends on a charge-weighted average of the volumes of each particle. 
}
Finally, combining the ion-exclusion correction terms, given in Eqs.~\ref{eqchargecavityii} and \ref{eqchargecavityij}, gives the predicted non-additivity of the cross interactions in a binary mixture in the DH limit as,
\begin{eqnarray}\label{eqnonaddDH}
\nonDH(\rr)   & = &    \frac{[\pot_{12}(\rr)]^{2}}{\pot_{11}(\rr) \pot_{22}(\rr)} - 1 \nonumber \\
			  & = &    \frac{[1 + \Omega_{12}]^{2}}{[1+\Omega_{11}] [1+\Omega_{22}]} - 1 
\end{eqnarray}
One sees immediately from Eqs.~\ref{eqchargecavityii} -- \ref{eqnonaddDH} that DLVO predictions are recovered ($\nonDH \rightarrow 0$) in the strongly screened limit, as $\invD \rr \rightarrow \infty$ at finite $\invD$, as one would naturally expect.

Our experiments examine the non-additivity of a size-asymmetric mixture in a limit where the particle pair separation is essentially fixed at $\rr = \rr^{*}$. To facilitate comparison between experiments and DH theory, we consider first the effect of size asymmetry on a mixture of charged spheres with equal reduced charges so,
\begin{equation}
\frac{Z_{1} \lB}{\collrad_{1}} = \frac{Z_{2} \lB }{\collrad_{2}}.
\end{equation}
In the limits of  extended double layers $\invD \collrad_{1} \ll 1$ and large  separations ($\rr \rightarrow \infty$), Eqs.~\ref{eqchargecavityii}--\ref{eqnonaddDH} reduce to a simple analytic expression for the non-additivity of mixtures with pure \textit{size asymmetry}, 
\begin{equation}\label{eqapproxasymmetricsize}
\nonDH(\invD) \approx - \frac{1}{3}(\invD \collrad_{1})^{3} (1-\gamma)^{2}(1+\gamma)  \frac{e^{-\invD \rr^{*} }}{\invD \rr^{*}}
\end{equation}
where the size ratio is $\gamma = \collrad_{2} / \collrad_{1}$ and the dependence of $\nonDH$ on the inverse screening length $\invD$ has been  noted explicitly.

\begin{figure}[h]
	%
	%
	\centering
	\includegraphics[trim = 10 10 330 30, width=0.45\textwidth]{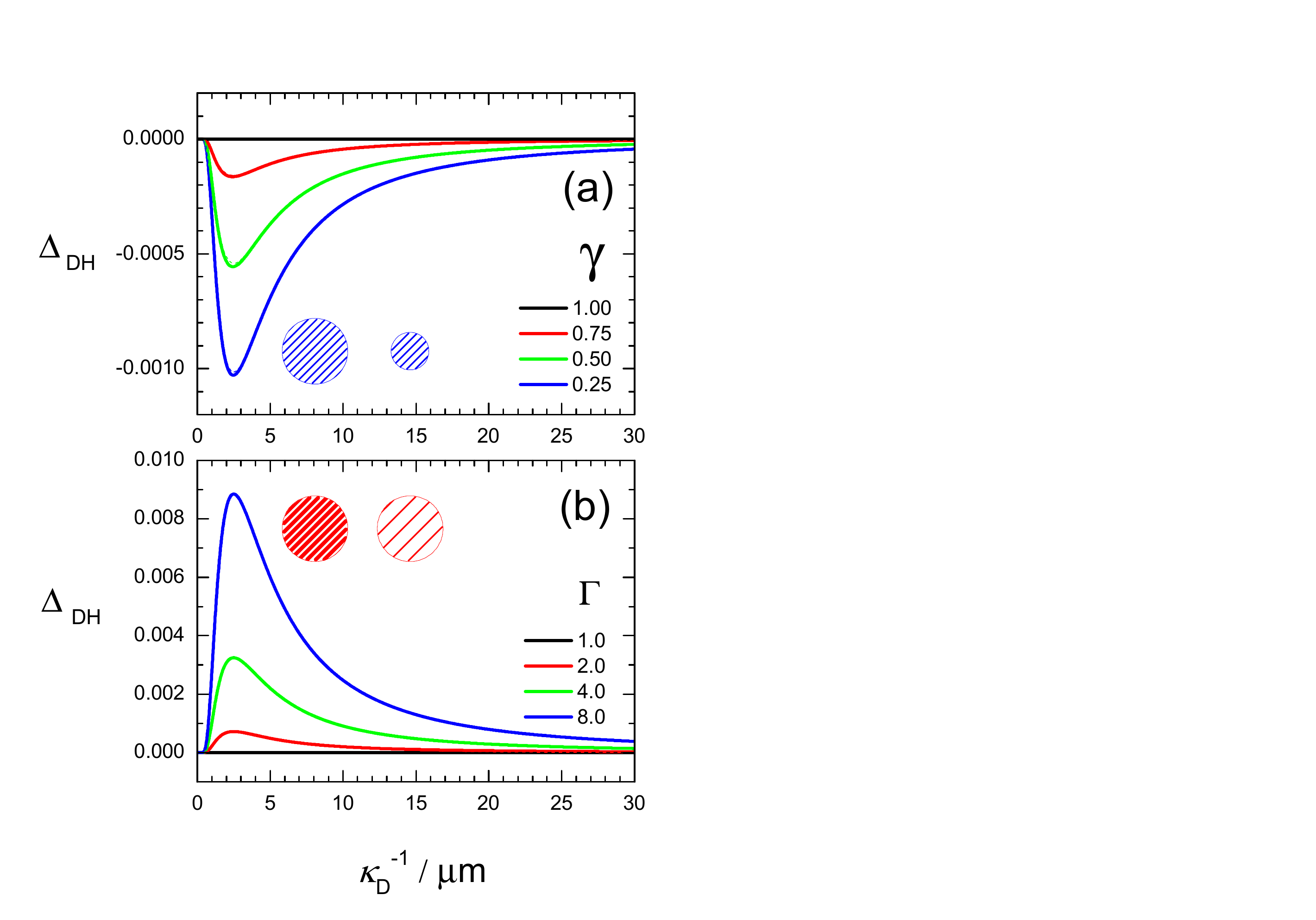}
	\caption{The non-additivity parameter $\nonDH$ calculated from  Debye-H\"{u}ckel theory for (a) a size-asymmetic mixture of equally-charged spheres of size ratio $\gamma=\collrad_{2} / \collrad_{1}$  and (b) a charge-asymmetric system of equally-sized spheres of charge ratio $\Gamma = Z_{2} / Z_{1}$. $\nonDH$ is plotted as a function of screening length $\invD^{-1}$, for $\rr^{*} = $ \SI{5}{\micro \meter} and $\collrad_{1} = $ \SI{1}{\micro \meter}. The solid lines in both graphs depict results from Eq.~\ref{eqnonaddDH} while the dashed lines in (a) and (b) (not visible beneath solid lines in (b)) show values from the approximate expressions, Eqs.~\protect\ref{eqapproxasymmetricsize} and ~\protect\ref{eqapproxasymmetriccharge}, respectively.}
	\label{fgrnonaddcalc}
\end{figure}

Figure~\ref{fgrnonaddcalc}(a) shows the high level of agreement between the approximation for the non-additivity in the size asymmetric system (Eq.~\ref{eqapproxasymmetricsize}, dashed lines) and the full expression  (Eq.~\ref{eqnonaddDH}, solid lines) at a fixed separation of $\rr^{*} = $ \SI{5}{\micro \meter}. Strikingly, we find a \textit{negative} value for the non-additivity, and a marked non-monotonic dependence on $\invD^{-1}$, with $\nonDH$ displaying a sharp minimum at $\invD \rr^{*} =2$ which grows in depth with increased size asymmetry. By contrast, same-sized mixtures with charge asymmetry {are predicted} to have \textit{positive} values of the  non-additivity, as illustrated by the data plotted in Fig.~\ref{fgrnonaddcalc}(b). From Eqs.~\ref{eqchargecavityii}--\ref{eqnonaddDH} we find, for $\invD \collrad \ll 1$ and $\rr \rightarrow \infty$, an analogous expression to Eq.~\ref{eqapproxasymmetricsize} in a system with pure \textit{charge asymmetry}, namely 
\begin{equation}\label{eqapproxasymmetriccharge}
\nonDH(\invD) \approx  \frac{1}{3}(\invD \collrad_{1})^{3} \frac{(\Gamma-1)^{2}}{\Gamma}  \frac{e^{-\invD \rr^{*} }}{\invD \rr^{*}}
\end{equation}
where $\Gamma = Z_{2} / Z_{1}$ is the charge ratio of species of equal radii $\collrad$. Overall, we see that analytical solutions of the Debye-H\"{u}ckel equations provide a simple conceptual picture for non-additivity in a binary mixture of charged colloids. Inclusion of the ion-cloud exclusion terms ensures that  the  cross-interaction is different from the geometric mean of the direct self interactions.  The sign of the corresponding non-additivity parameter $\nonDH$ depends sensitively on the degree of both charge and size asymmetry. Mixtures of spheres with the same surface charge but different sizes display negative non-additivity parameters whilst equal-sized mixtures with different charges display positive values of non-additivity.

To compare the DH predictions with experiments we assume that the linear screening theory is qualitatively correct in predicting the sign of the non-additivity and semi-quantitatively predicts the functional dependence of $\nonadd$ on the screening length $\invD^{-1}$. We therefore utilize linear screening theory (Eqs.~\ref{eqchargecavityii}--\ref{eqnonaddDH}) to find $\nonDH$ and substitute these values into the expression for the measured non-additivity, $\nonadd(\invD) = C \nonDH(\invD)$, where $C$ is a scaling constant. The full lines in Fig.~\ref{fgrnonaddall} show the resulting fits to the experimental non-additivities, where the scaling constant $C$ and the fixed pair separation $\rr^{*}$ have been used as fit parameters. We obtain $\rr^{*} = 10 \pm 1$ \si{\micro \meter}, a scale constant of $C= 900 \pm 300$ at $\gamma = 0.59$, and a minimum non-additivity parameter of $\nonadd = -0.09$. In a similar manner we calculate the fitting parameters at $\gamma = 0.40$ as  $\rr^{*} = 4.0 \pm 0.3$ \si{\micro \meter}, a scale constant of $C= 44 \pm 8$, and a minimum non-additivity parameter of $\nonadd = -0.14$. The dependence of the measured non-additivity on the screening length is seen to follow reasonably well the same trend as the DH expression which suggests that the origin of the non-additivity is indeed the doubly-screened ion exclusion terms. However the large discrepancy in magnitude between the predictions and experiments implies that the strength of ion exclusions terms are significantly underestimated by DH  theory, which is probably not too surprising as the experimental particles charges lie firmly in the nonlinear regime where $Z \lB / \collrad_{1} \gg 1$. Presumably the high ion densities near the surface of the particle in nonlinear PB theory enhance the strength of the ion exclusion terms.  { We are not aware of any direct single-particle measurements which corroborate the predictions of positive non-additivity in mixtures of equal-sized spheres with different charges. Experiments are planned in the near future.}

Finally, we explore the significance of non-additivity for the phase behaviour of binary charged suspensions, by mapping the Yukawa system onto a binary hard sphere (HS) mixture. Introducing the Barker-Henderson effective hard core diameters $\sigma_{ij} = \int_{0}^{\infty} d\rr \left\{ 1 - \exp[-\pot_{ij}(\rr) / \kBT ]  \right\}$ where $(i,j)= 1,2$ then the equivalent hard-sphere non-additivity parameter is { $\Delta_{\mys{hs}}$ where $\sigma_{12} = \frac{1}{2}(\sigma_{11}+\sigma_{22}) (1+\Delta_{\mys{hs}})$.} Evaluating this expression numerically, we find that $\Delta_{\mys{hs}} = -0.02$ and $-0.03$ for the charged systems at radius ratios $\gamma = 0.59$ and 0.40, respectively. The influence of non-additivity on the global topology of HS phase diagrams has been widely studied\cite{Louis2000,Santos2005a}. The existence of a fluid-fluid phase separation in HS mixtures has been shown, for instance, to be strongly influenced by even modest degrees of non-additivity\cite{Louis2000}. For a sufficiently negative $\Delta_{\mys{hs}}$ local clustering and the formation of heterogeneous structures\cite{Hoffmann2006} are observed in which small particles tend to occupy the voids between big particles, similar to the chemical short-range order observed  in amorphous and liquid binary mixtures with preferred hetero-coordination.  { Negative non-additive or purely additive binary Yukawa systems are not expected to phase separate\cite{Hopkins2006} but a fluid-fluid demixing transition is generated by even relatively small degrees\cite{Louis2000,Pellicane2006} of positive non-additivity ($\Delta_{\mys{hs}} > 0.01$). Consequently, we suspect that the positive non-additivity predicted for equal-sized mixtures with different charge levels may be strong enough to drive a fluid-fluid phase separation.}  Interestingly, such a phase separation has been reported in charge asymmetric mixtures\cite{Yoshizawa2012}, precisely under conditions where we predict a small positive non-additivity. In addition, non-additivity effects are expected to have a particularly pronounced effect on the stability of mixed crystalline phases\cite{Barrat1992}. Indeed it is possible that assuming $\nonadd \neq 0$  might account for the puzzling observation of non-close-packed phase formation in binary nanocrystal mixtures\cite{Shevchenko2006}, which is inconsistent\cite{Bodnarchuk2010,Boles2015} with calculations where the non-additivity parameter has been set to zero.

\section{Conclusions}

Our results provide direct experimental evidence for the emergence of significant non-additivity effects in the pair interactions of binary charged microparticles. We confirm that the non-additivity parameter $\nonadd$ in size-asymmetric Yukawa mixtures is negative. Within Debye-H\"{u}ckel theory, we propose a simple conceptual framework for $\nonadd$ and show how the magnitude and sign of the non-additivity can be tuned, for example, by altering the screening length, the size ratio, or the charge ratio of the two particles. 
We anticipate that these findings may be useful to rationalize the complex phase behaviour seen in binary colloidal\cite{Yoshizawa2012,Lorenz2009} and nanoparticle\cite{Shevchenko2006} systems as well as providing design rules for non-additive potentials suitable for pre-programmed self-assembly\cite{Salgado-Blanco2015}.

\section*{Acknowledgments}
We thank D Gillespie, G Smith, and I Williams for help. The work was part supported by the EPSRC via a Doctoral Training Award to S.D.F.

%


\end{document}